%% file: main.tex
\documentclass[journal,twoside]{IEEEtran}

\usepackage[utf8]{inputenc} 
\usepackage[T1]{fontenc}
\usepackage{url}
\usepackage{ifthen}
\usepackage{graphicx}
\usepackage[T1]{fontenc}

\usepackage{enumitem}
\usepackage[cmex10]{amsmath}
\usepackage{amssymb}

\usepackage{adjustbox}
\usepackage{cite}
\usepackage{amsthm}
\usepackage{textcomp}
\usepackage{siunitx}
\usepackage{color}
\usepackage{cases}
\usepackage[font={small}]{caption}
\usepackage{epstopdf}
\usepackage{graphicx}
\usepackage{caption}
\usepackage{verbatim} 
\usepackage{bbm}
\usepackage{mathtools}

\newtheorem{theorem}{Theorem}
\newtheorem{lemma}{Lemma}

\newtheorem{remark}{Remark}
\newtheorem*{theorem*}{Theorem}

\theoremstyle{definition}
\newtheorem{example}{Example}
\newtheorem{definition}{Definition}

\theoremstyle{remark}

\setlist[description]{style=multiline}

\begin{document}
\sloppy

\title{Coded Computing for Secure Boolean Computations} 

\author{Chien-Sheng Yang,~\IEEEmembership{Graduate Student Member,~IEEE}, and A. Salman Avestimehr,~\IEEEmembership{Fellow,~IEEE}

\thanks{This material is based upon work supported by Defense Advanced Research Projects Agency (DARPA) under Contract No. HR001117C0053, ARO award W911NF1810400, NSF grants CCF-1703575 and CCF-1763673,  and ONR Award No. N00014-16-1-2189. The views, opinions, and/or findings expressed are those of the author(s) and should not be interpreted as representing the official views or policies of the Department of Defense or the U.S. Government. A part of this paper was presented in ISITA2020\cite{Yang2010:Coded}.
}
\thanks{C.-S.~Yang and A.~S.~Avestimehr are with the Department of Electrical and Computer Engineering, University of Southern California, Los Angeles, CA 90089 USA (e-mail: chienshy@usc.edu; avestimehr@ee.usc.edu).}

}
\maketitle

\begin{abstract}
\input{0-abstract}
\end{abstract}
\begin{IEEEkeywords}
Boolean Function, Coded Computing, Distributed computing
\end{IEEEkeywords}
\section{Introduction}
\input{1-intro}

\section{System Model}\label{sec:sys}
\input{2-system}
\section{Overview of Lagrange Coded Computing}\label{sec:lcc}
\input{3-LCC}
\section{Scheme 1: Coded Algebraic Normal Form}
\input{4-ANF}
\section{Scheme 2: Coded Disjunctive Normal Form}
\input{5-DNF}
\section{Scheme 3: Coded Polynomial Threshold Function}\label{sec:PTF}
\input{6-PTF}
\section{Matching Outer Bound for coded ANF and coded DNF}
\input{7-lowerbound}
\section{Application to Cryptography}
\input{8-example}

\section{Extension to General multivariate  Polynomials}
\input{9-sparse_poly}
\section{Concluding Remarks and Future Directions}
\input{10-conclusion}
\bibliographystyle{ieeetr}
\bibliography{references}
\vspace{-5mm}
\begin{IEEEbiographynophoto}{Chien-Sheng Yang}
received his the B.S. degree in electrical and computer engineering from
National Chiao Tung University (NCTU), Hsinchu, Taiwan in 2015 and is currently pursuing his Ph.D. in Electrical and Computer Engineering from the University of Southern California (USC), Los Angeles. He received the Annenberg Graduate Fellowship in 2016. He was a finalist of the ACM International Symposium on Mobile Ad Hoc Networking and Computing (MobiHoc) Best Paper Award in 2019. His interests include information theory, machine learning and edge computing. 
\end{IEEEbiographynophoto}
\vspace{-5mm}
\begin{IEEEbiographynophoto}{A. Salman Avestimehr}
is a Professor and director of the Information Theory and Machine Learning (vITAL) research lab at the Electrical and Computer Engineering Department of University of Southern California. He is also an Amazon Scholar at Alexa AI. He received his Ph.D. in 2008 and M.S. degree in 2005 in Electrical Engineering and Computer Science, both from the University of California, Berkeley. Prior to that, he obtained his B.S. in Electrical Engineering from Sharif University of Technology in 2003. His research interests include information theory and coding theory, and large-scale distributed computing and machine learning, secure and private computing, and blockchain systems

Dr. Avestimehr has received a number of awards for his research, including the James L. Massey Research $\&$ Teaching Award from IEEE Information Theory Society, an Information Theory Society and Communication Society Joint Paper Award, a Presidential Early Career Award for Scientists and Engineers (PECASE) from the White House, a Young Investigator Program (YIP) award from the U.S. Air Force Office of Scientific Research, a National Science Foundation CAREER award, the David J. Sakrison Memorial Prize, and several Best Paper Awards at Conferences. He has been an Associate Editor for IEEE Transactions on Information Theory. He is currently a general Co-Chair of the 2020 International Symposium on Information Theory (ISIT).
\end{IEEEbiographynophoto}
\clearpage
\appendices
\section{Proof of Theorem 5}
\input{Appendix-proof}

\end{document}

%% file: 0-abstract.tex
The growing size of modern datasets necessitates splitting a large scale computation into smaller computations and operate in a distributed manner. Adversaries in a distributed system deliberately send erroneous data in order to affect the computation for their benefit. Boolean functions are the key components of many applications, e.g., verification functions in blockchain systems and design of cryptographic algorithms. We consider the problem of computing a Boolean function in a distributed computing system with particular focus on \emph{security against Byzantine workers}. Any Boolean function can be modeled as a multivariate polynomial with high degree in general. However, the security threshold (i.e., the maximum number of adversarial workers can be tolerated such that the correct results can be obtained) provided by the recent proposed Lagrange Coded Computing (LCC) can be extremely low if the degree of the polynomial is high. We propose three different schemes called \emph{coded Algebraic normal form (ANF)}, \emph{coded Disjunctive normal form (DNF)} and \emph{coded polynomial threshold function (PTF)}. The key idea of the proposed schemes is to model it as the concatenation of some low-degree polynomials and threshold functions. In terms of the security threshold, we show that the proposed coded ANF and coded DNF are optimal by providing a matching outer bound. 



%% file: 1-intro.tex
With the growing size of modern datasets for applications such as machine learning and data science, it is necessary to partition
a massive computation into smaller computations and perform these smaller computations in a distributed manner for improving overall performance~\cite{abadi2016tensorflow}. However, distributing the computations to some external entities, which are not necessarily trusted, i.e., adversarial servers make security a major concern~\cite{blanchard2017machine,cramer2015secure,bogdanov2008sharemind}.  
Thus, it is important to provide security against adversarial workers that deliberately send erroneous data in order to affect the computation for their benefit.

Boolean functions are primarily used in the design of cryptographic algorithms~\cite{cusick2017cryptographic}. In particular, computing Boolean functions is one of the key components of blockchains. In the blockchain systems, Boolean functions can be used to represent the verification functions which validate the transactions in the new proposed blocks~\cite{9141331}. Specifically, each node computes function \texttt{is}$\_$\texttt{valid}$\_$\texttt{txn}$\in\{$\texttt{True,False}$\}$ to determine whether a transaction is valid or not~\cite{cao2020cover}. Due to the heavy computation cost incurred by validating all the blocks, the nodes with limited resources cannot verify all the blocks independently. To improve the efficiency (e.g., number of transactions verified by the system), the leading solution is via sharding~\cite{luu2016secure} whose idea is to partition the blockchain into sub-chains and the block validations are executed distributively in each node.



In this paper, we consider the problem of computing a Boolean function (e.g., block validation) in which the computation is carried out distributively across several workers with particular focus on \emph{security against Byzantine workers}. Specifically, using a master-worker distributed computing system with $N$ workers, the goal is to compute the Boolean function $f: \{0,1\}^m \rightarrow \{0,1\}$ over a dataset of $K$ samples $X_1, \dots, X_K$, i.e., $f(X_1),\dots,f(X_K)$, in which the (encoded) datasets are prestored in the workers such that the computations can be secure against adversarial workers in the system. Especially, we consider the adversarial model in which the malicious workers do not have any computational restriction and are capable of sending erroneous data. To measure the robustness against adversaries of a given scheme $S$, we use the metric \emph{security threshold} $\beta_{S}$ which is defined as the maximum number of adversarial workers that can be tolerated by the master, i.e., the correct results can be recovered even if there are up to $\beta_S$ adversarial workers.

Any Boolean function can be modeled as an Algebraic normal form (i.e., multivariate polynomial)~\cite{cusick2017cryptographic}. Thus, the recently proposed Lagrange Coded Computing (LCC) \cite{yu2019lagrange}, a universal encoding technique for arbitrary multivariate polynomial computations, can be used to simultaneously alleviate the issues of resiliency, security, and privacy. In overview, for the problem of computing an arbitrary multivariate polynomial $f: \mathbb{V} \rightarrow \mathbb{U}$ over a field $\mathbb{F}$, LCC encodes $X_1,\dots,X_K \in \mathbb{V}$ by evaluating the well-known Lagrange polynomial, and each encoded data is stored in a different worker. The workers then apply the multivariate polynomial of interest $f$ (e.g., Boolean function) on their encoded data and return the computation results back to the master. Since the computation executed in each worker can be viewed as a composition of a multivariate polynomial and a univariate polynomial, the problem becomes a polynomial interpolation with errors and erasures. The master recovers the computation by evaluating the interpolated polynomial at the appropriately chosen points. 

The security threshold provided by LCC is $\bigl \lfloor \frac{N - (K-1)\textrm{deg} f -1}{2} \bigr \rfloor$ (given $N$ and $K$) which can be extremely low if the degree of corresponding multivariate polynomial $\textrm{deg} f$ is high (see more details in Section \ref{sec:lcc}). Such degree problem can be further amplified in complex Boolean functions whose degree can be high in general. Thus, our main problem is as follows: What is the maximum possible security threshold and the corresponding scheme, given $f$, $N$ and $K$? 
\subsection{Main Contributions}
\begin{figure}[t]
    \centering
    \includegraphics[width = 0.75\columnwidth]{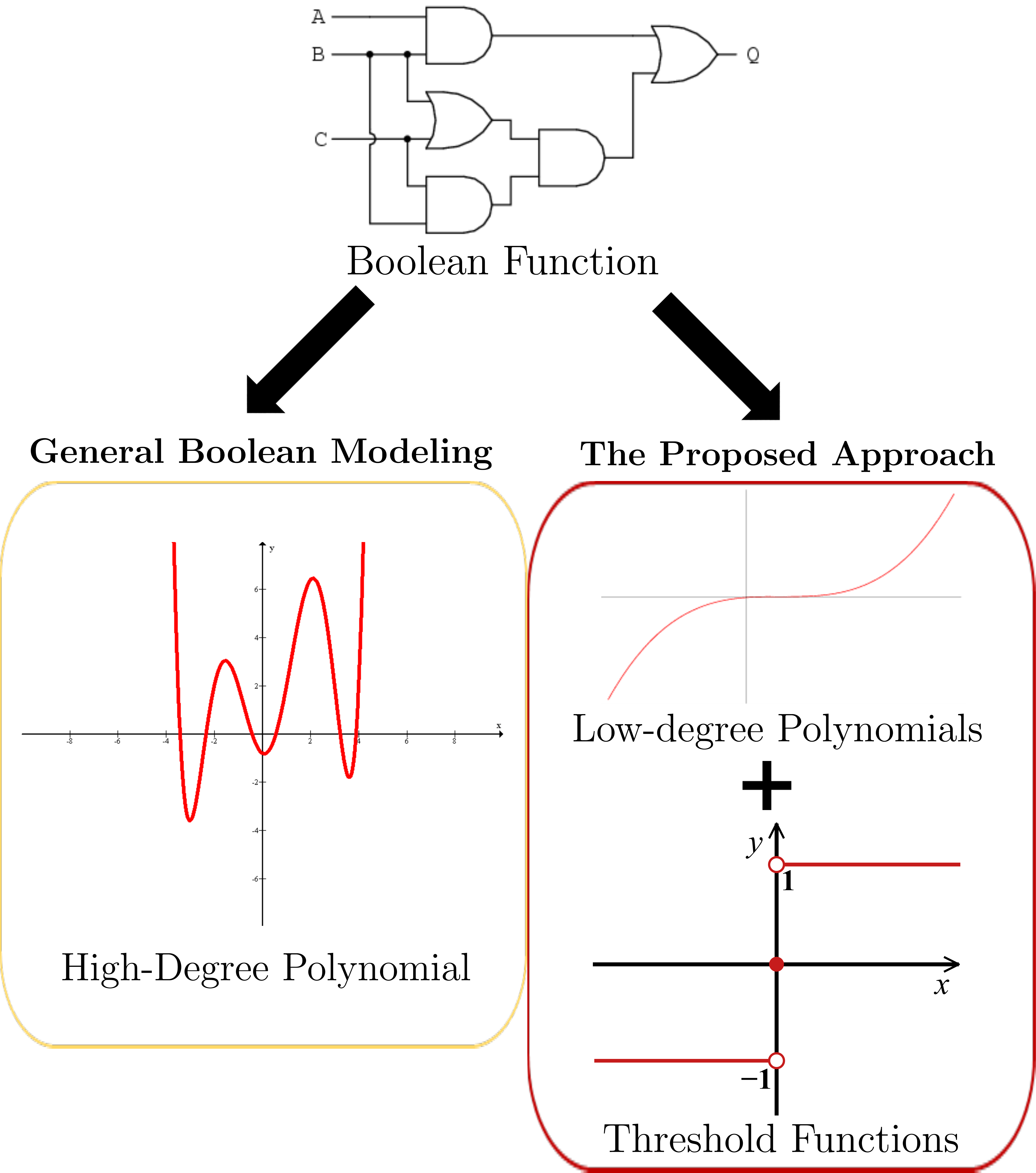}
    \caption{Modeling the Boolean function as a general polynomial can result in the high-degree difficulty which makes the security threshold low by using LCC encoding. The main idea of our proposed approach is to model it as the concatenation of some low-degree polynomials and the threshold functions.
   } 
    \label{fig:boolean}
\end{figure}
As main contributions of the paper, instead of modeling the Boolean function as a general polynomial, we propose the three schemes modeling it as the concatenation of some low-degree polynomials and the threshold functions (see Figure \ref{fig:boolean}). 
To illustrate the main idea of the proposed schemes, consider an AND function of three input bits $X[1],X[2],X[3]$ which is formally defined by $f(X) = X[1] \wedge X[2] \wedge X[3]$. The function $f$ can be modeled as a polynomial function (Algebraic normal form) $X[1]X[2]X[3]$ which has a degree of $3$. For this polynomial, LCC achieves the security threshold $\bigl\lfloor\frac{N-3(K-1)-1}{2}\bigr\rfloor$. Instead of directly computing the degree-$3$ polynomial, our proposed approach is to model it as a linear threshold function $\text{sgn}(X[1]+X[2]+X[3]-\frac{5}{2})$ in which $f(X) = 1$ if and only if $\text{sgn}(X[1]+X[2]+X[3]-\frac{5}{2}) > 0$. Then, a simple linear code (e.g., $(N,K)$ MDS code) can be used for computing the linear function $X[1]+X[2]+X[3]$, which provides the optimal security threshold $\bigl\lfloor \frac{N-K}{2} \bigr\rfloor$.

We propose three different schemes called \emph{coded Algebraic normal form (ANF)}, \emph{coded Disjunctive normal form (DNF)} and \emph{coded polynomial threshold function (PTF)}. The idea behind coded ANF (DNF) is to first decompose the Boolean function into some monomials (clauses) and then construct a linear threshold function for each monomial (clause). For both of coded ANF and coded DNF, an $(N,K)$ MDS code is used to encode the datasets. On the other hand, the proposed coded PTF models the Boolean function as a low-degree polynomial threshold function, and LCC is used for the data encoding. 

 For any general Boolean function $f$, the proposed coded ANF and coded DNF achieve the security threshold $\bigl\lfloor \frac{N-K}{2} \bigr\rfloor$, which is independent of $\textrm{deg} f$. In terms of security threshold, we prove that coded ANF and coded DNF are optimal by deriving a matching theoretical outer bound. To demonstrate the impact of coded ANF and coded DNF, we consider the problem of computing $8$-bit S-box in the application of block cyphers using a distributed computing system with $100$ workers. We show that coded ANF and coded DNF can significantly improve the security threshold by $150\%$ as compared to LCC.

 In Table \ref{table:compare_all}, we summarize the performance comparison of LCC and the proposed three schemes in terms of the security threshold and the decoding complexity. As compared to LCC, coded ANF and coded DNF provide the substantial improvement on the security threshold. In particular, coded ANF has the decoding complexity $O(r(f)N \log^2{N} \log{\log{N}})$ which works well for the Boolean functions with low sparsity $r(f)$; coded DNF has the decoding complexity $O(w(f)N \log^2{N} \log{\log{N}})$ which works well for the Boolean functions with small weight $w(f)$ (see the definitions of $r(f)$ and $w(f)$ in Section \ref{sec:sys}). For the Boolean functions with the polynomial size of $r(f)$ and $w(f)$, coded PTF outperforms LCC by achieving the better security threshold and the almost linear decoding complexity which is independent of $m$ (see more details in Section~\ref{sec:PTF}). 
 
 Finally, We extend the problem to a more general computation model, i.e., $f$ is a multivariate polynomial function. To resolve the high-degree difficulty arising in computing general polynomials, we propose two schemes: \textit{coded data logarithm} and \textit{coded data augmentation}. By taking the logarithm of original data, the proposed coded data logarithm scheme reduces the degree of polynomial computations, and improves the security threshold as compared to LCC. On the other hand, the proposed coded data augmentation scheme pre-stores some low-degree monomials in advance to make the polynomial computation's degree reduced.
\begin{table}[t]
  \centering
  \scalebox{0.85}{
  \begin{tabular}{| c | c | c | }
    \hline
    \rule{0pt}{12pt} & \textbf{Security Threshold} & \textbf{Decoding Complexity} \\ \hline
    \rule{0pt}{12pt} \textbf{LCC}  & $\bigl\lfloor \frac{N - (K-1)\textrm{def} f -1}{2} \bigr\rfloor$  &  $O(mN\log^3{N} \log{\log{N}})$ \\ [1ex] \hline
    \rule{0pt}{12pt}  \textbf{Coded ANF} &  $\bigl\lfloor \frac{N-K}{2} \bigr\rfloor$  & $O(r(f)N \log^2{N} \log{\log{N}})$ \\ [1ex] \hline
     \rule{0pt}{12pt} \textbf{Coded DNF}  & $\bigl\lfloor \frac{N-K}{2} \bigr\rfloor$  & $O(w(f)N \log^2{N} \log{\log{N}})$ \\ [1ex] \hline
     \rule{0pt}{12pt} \textbf{Coded PTF} & $\bigl\lfloor \frac{N - (K-1)(\lfloor \log_2{w(f)} \rfloor+1) -1}{2} \bigr\rfloor$  &$O(N\log^2{N} \log{\log{N}})$ \\ [1ex] \hline
     \rule{0pt}{12pt} \textbf{Outer Bound} & $\bigl\lfloor \frac{N-K}{2} \bigr\rfloor$ & -\\ [1ex] \hline
  \end{tabular}
  }
  \caption{Performance comparison of LCC and the proposed three schemes for the Boolean function $f:\{0,1\}^m \rightarrow \{0,1\}$ which has the sparsity $r(f)$ and weight $w(f)$. }\label{table:compare_all}
\end{table}

\subsection{Related Prior Work}
Next, we provide a brief literature review that covers two main lines of work: polynomial threshold functions representing Boolean functions, and coded computing. 

The expressive power of real polynomial threshold functions for representing Boolean functions has been extensively studied over the decades. The study of representing Boolean functions by polynomial threshold functions was initiated in~\cite{rosenblatt1958perceptron,block1962perceptron,minsky1969perceptrons}. The following works focused largely on the degree of PTF needed to represent a Boolean function (e.g.,~\cite{nisan1994degree,aspnes1994expressive,o2003new,o2008extremal,bun2019nearly}), and the density of PTF needed to represent a Boolean function (e.g,~\cite{oztop2006upper,o2008extremal,amano2010new,sezener2015heuristic}). Polynomials threshold functions also play a vital role in complexity theory and learning theory (e.g.,~\cite{klivans2004learning2,alman2016polynomial}).

Coded computing broadly refers to a family of techniques that utilize coding to inject computation redundancy in order to alleviate the various issues that arise in large-scale distributed computing. In the past few years, coded computing has had a tremendous success in various problems, such as straggler mitigation and bandwidth reduction (e.g., \cite{lee2018speeding,li2018fundamental,dutta2016short,lee2017high,yu2017polynomial,tandon2017gradient,li2017coding,narra2019slack,li2020coded,prakash2020codedgraph,guler2020tacc,yu2020straggler}). Coded computing has also been expanded in various directions, such as heterogeneous networks (e.g.,~\cite{reisizadeh2019coded}), partial stragglers (e.g.,~\cite{ferdinand2018hierarchical}), secure and private computing (e.g.,~\cite{chen2018draco,yu2019lagrange,kadhe2019gradient,nodehi2019secure,so2019codedprivateml,so2020scalable,yu2020coded,soleymani2020analog}), distributed optimization (e.g.,~\cite{karakus2017straggler}), federated learning (e.g.,~\cite{so2020turbo,prakash2020hierarchical,prakash2020coded}), blockchains (e.g.,~\cite{yu2020codedtree,9141331}) and dynamic networks (e.g.,~\cite{yang2019timelyisit,yang2019timely,yang2021edge}).

So far, research in coded computing has focused on developing frameworks for some linear functions (e.g., matrix multiplications). However, there has been no works prior to our work that consider coded computing for Boolean functions. In this paper, we make the substantial progress of improving the security threshold by proposing coded ANF, coded DNF and coded PTF which leverage the idea of the threshold function representation.

\textbf{Notation.} For the Boolean logical operations, we denote the logical operators of AND, OR, XOR and NOT by $\wedge$, $\vee$, $\oplus$ and $\sim$ respectively.

%% file: 2-system.tex
We consider the problem of evaluating a Boolean function $f: \{0,1\}^m \rightarrow \{0,1\}$ over a dataset $\vec{X} = (X_1,\dots,X_K)$, where $X_1,\dots,X_K$ are $m$-dimensional vectors over the field $\{0,1\}$. Given a distributed computing environment with a master and $N$ workers, our goal is to compute $f(X_1), \dots, f(X_K)$.  

Each Boolean function $f: \{0,1\}^m \rightarrow \{0,1\}$ can be represented by an Algebraic normal form (ANF)~\cite{cusick2017cryptographic,o2014analysis} as follows:
\begin{align}
    f(X) = \bigoplus_{\mathcal{S} \subseteq [m]} \mu_f(\mathcal{S}) \prod_{j \in \mathcal{S}} X[j] \label{eq:ANF}
\end{align}
where $X[j]$ is the $j$-bit of data $X$ and $\mu_f(\mathcal{S}) \in \{0,1\}$ is the ANF coefficient of the corresponding monomial $\prod_{j \in \mathcal{S}} X[j]$. The total degree\footnote{The total degree of a multivariate polynomial is the maximum among all the total degrees of its monomials.} of the ANF representation of Boolean function $f$ is denoted by $\textrm{deg} f$. We denote the sparsity (number of monomials) of $f$ by $r(f)$, i.e., $r(f) = \sum_{\mathcal{S} \subseteq [m]}\mu_f(S)$. Since each monomial in ANF has the degree up to $\textrm{deg}f$, the total complexity of computing $f(X_1),\dots,f(X_K)$ via ANF of $f$ is $O(Kr(f)\textrm{deg}f)$.

Furthermore, we denote the support of $f$ by $\textrm{Supp}(f)$ which is the set of vectors in $\{0,1\}^m$ such that $f=1$, i.e., $\textrm{Supp}(f) = \{X \in \{0,1\}^m: f(X) = 1\}$. Let $w(f)$ be the weight of Boolean function $f$, defined by $w(f) =|\textrm{Supp}(f)|$. Alternatively, each Boolean function $f$ can be represented by a Disjunctive normal form (DNF) as follows:
\begin{align}
    f=T_1 \vee T_2 \vee \dots \vee T_{w(f)} \label{eq:DNF}
\end{align}
where each clause $T_i$ has $m$ literals\footnote{A literal is a Boolean variable or the complement of a Boolean variable.} in which each literal corresponds to an input $Y_i$ such that $f(Y_i) = 1$. For example, if $Y_i = 001$, then the corresponding clause is $\sim Y_i[1]\wedge \sim Y_i[2] \wedge Y_i[3]$. Since each clause of DNF has $m$ literals, the total complexity of computing $f(X_1),\dots,f(X_K)$ via DNF of $f$ is $O(Kmw(f))$.

Prior to computation, each worker has already stored a fraction of the dataset in a possibly coded manner. Specifically, each worker $n$ stores $\tilde{X}_n = g_n(X_1,\dots,X_K)$, where $g_n:\underbrace{\{0,1\}^m \times \cdots \times \{0,1\}^m}_{K} \rightarrow \mathbb{U}$ is the encoding function of worker $n$ and $\mathbb{U}$ is an arbitrary vector space. We restrict our attention to linear encoding schemes, which guarantee low encoding complexity. Each worker $n$ computes $h(\tilde{X}_n)$ and returns the result back to the master, in which $h$ is the  the multivariate polynomial function decided by the master and $f(X)$ is function of $h(X)$. Then, the master aggregates the results from the workers until it receives a \emph{decodable} set of local computations. We say a set of computations is decodable if $h(X_1),\dots,h(X_K)$ can be obtained by computing decoding functions over the received results.

More concretely, given any subset of workers that return the computing results (denoted by $\mathcal{K}$), the master computes $v_{\mathcal{K}}(\{h(\tilde{X}_n)\}_{n \in \mathcal{K}})$, where each $v_{\mathcal{K}}$ is a deterministic function. We refer to the $v_{\mathcal{K}}$’s as decoding functions. Finally, the master computes $f(X_1),\dots,f(X_K)$ based on $h(X_1),\dots,h(X_K)$.

In particular, we focus on finding the scheme $(\vec{g},h)$ to be robust to as many adversarial workers as possible in the system where $\vec{g} = (g_1,\dots,g_N)$ is the collection of encoding functions. To measure the robustness against adversaries of a given scheme, we use the metric \emph{security threshold} defined as follows:
\begin{definition}[Security Threshold] \label{def:security}
For an integer $b$, we say a scheme $S$ is $b$-secure if the master can be robust against $b$ adversaries, i.e., the master can recover all the correct results even if up to $b$ workers return arbitrarily erroneous results. The security threshold, denoted by $\beta_S$, is the maximum value of $b$ such that a scheme $S$ is $b$-secure, i.e.,
\begin{align}
    \beta_S \triangleq \sup \{b:S \ \text{is} \ b\text{-secure}\}.
\end{align}
\end{definition}
Based on the above system model, the problem is now formulated as: \emph{What is the scheme which achieves the optimal security threshold with low decoding complexity?}
\begin{remark}
To see how much computation cost that the master can save using a given scheme, it is important to compare the total complexity of computing $K$ evaluations $f(X_1),\dots,f(X_K)$ (by the master itself) with the complexity incurred by the scheme. Since the encoding process of a scheme is only executed once before starting any computations, we focus on the decoding complexity which is the main cost incurred by a scheme throughout the paper. 
\end{remark}

\begin{remark}
    To see how the distributed Boolean computation is applicable to a sharded blockchain system, we can consider a blockchain system $\texttt{PolyShard}$ [7] which is implemented distributedly over some untrusted nodes. At each time epoch, each node stores a coded version of sub-chain and computes a validation function directly on the coded sub-chain and a coded block (generated by computing an encoding function on the incoming blocks). After the computations, each node broadcasts the computed result to all other nodes. Then, each node computes the decoding function on the received computation results to reduce the desired validation result and determines the validity of block. That is, each node plays the role of a master node after the procedure of broadcasting. When there is a new participant joining the network, a new coded sub-chain can be generated and stored in this new node. When there is a participant leaving the network, the blockchain with remaining nodes can still work since each node stores a coded sub-chain and the system can follow the same procedure for the block validations.
    \end{remark}

%% file: 3-LCC.tex
In this section, we consider the recently proposed Lagrange Coded Computing (LCC) \cite{yu2019lagrange}, which is a universal encoding technique for the class of multivariate polynomial functions. Then, we show how it works for our problem. 

Since Lagrange coded computing requires the underlying field size to be at least the number of workers $N$, we first extend the field size of $\{0,1\}$ such that the size of extension field is at least the number of workers $N$. More specifically, we embed each bit $X_k[j] \in \{0,1\}$ of data $X_k$ into a binary extension field $\mathbb{F}_{2^t}$ such that with $2^t \geq N$. The embedding $\bar{X}_k[j] \in \mathbb{F}_{2^t}$ of the bit $X_k[j]$ is generated such that 
\begin{align}
    \bar{X}_k[j] = 
    \begin{cases}
    \underbrace{00\cdots0}_{t}, \ X_k[j] = 0, \\
    \underbrace{00\cdots0}_{t-1}1, \ X_k[j] = 1.
    \end{cases}
\end{align}
Note that over extension field the output of Boolean function $f$ is $ \underbrace{00\cdots0}_{t}$ if the original result is $0$; $\underbrace{00\cdots0}_{t-1}1$ if the original result is $1$.

For the data encoding by using LCC, we first select $K$ distinct elements $\beta_1, \beta_2,\dots, \beta_K$ from the binary extension field $\mathbb{F}_{2^t}$, and let $u$ be the respective \emph{Lagrange interpolation polynomial}: 
\begin{align}
    u(z) \triangleq \sum^{K}_{k=1}\bar{X}_k \prod_{l \in [K]\backslash \{k\}}\frac{z-\beta_l}{\beta_k - \beta_l},
\end{align}
where $u: \mathbb{F}_{2^t} \rightarrow \mathbb{F}^m_{2^t}$ is a polynomial of degree $K-1$ such that $u(\beta_k) = \bar{X}_k$. Then we can select distinct elements $\alpha_1,\alpha_2,\dots,\alpha_N \in \mathbb{F}_{2^t}$, and encode $\bar{X}_1,\dots,\bar{X}_K$ to $\tilde{X}_n = u(\alpha_n)$ for all $n \in [N]$, i.e., 
\begin{align}
    \tilde{X}_n = u(\alpha_n) \triangleq  \sum^{K}_{k=1}\bar{X}_k \prod_{l \in [K]\backslash \{k\}}\frac{\alpha_n-\beta_l}{\beta_k - \beta_l}.
\end{align}
Each worker $n \in [N]$ stores $\tilde{X}_{n}$ locally. Following the above data encoding, each worker $n$ computes function $f$ on $\tilde{X}_n$ and sends the result back to the master upon its completion. Since the computation is over the extension field, the complexity at each worker is $O(tr(f)\textrm{deg}f)$.

After receiving results from all the workers, the master can obtain all coefficients of $f(u(z))$ by applying Reed-Solomon decoding \cite{berlekamp1968nonbinary,massey1969shift}. Having this polynomial, the master evaluates it at $\beta_k$ for every $k \in [K]$ to obtain $f(u(\beta_k)) = f(\bar{X}_k)$. The complexity of decoding a length-$N$ Reed-Solomon code with dimension $(K-1)\textrm{deg}f + 1$ for one symbol over the extension field is $O(tN \log^2{N\log{\log{N}}})$. 
To have a sufficiently large field for LCC, we pick $t = \lceil \log{N}\rceil$. Since there are $m$ symbols in each $\tilde{X}_n$, the decoding process by the master requires complexity $O(mN\log^3{N} \log{\log{N}})$.

In the following, we present the security threshold provided by LCC. By \cite{yu2019lagrange}, to be robust to $b$ adversarial workers (given $N$ and $K$), LCC requires $N\geq (K-1)\textrm{deg} f +2b+1$; i.e., LCC achieves the security threshold
\begin{align}
    \beta_{\textrm{LCC}} = \bigl \lfloor \frac{N - (K-1)\textrm{deg} f -1}{2} \bigr \rfloor.
\end{align}
 
The security threshold achieved by LCC depends on the degree of function $f$, i.e., the security guarantee is highly degraded if $f$ has high degree. To mitigate such degree effect, we model the Boolean function as the concatenation of some low-degree polynomials and the threshold functions by proposing three schemes in the following sections. 

%% file: 4-ANF.tex
In this section, we propose a coding scheme called \emph{coded Algebraic normal form (ANF)} which computes the ANF representations of Boolean function by the linear threshold functions (LTF) and a simple linear code is used for the data encoding. We start with an example to illustrate the idea of coded ANF.
\begin{example}
We consider a function which has an ANF representation defined as follows:
\begin{align}
    f(X) = X[1]X[2]\cdot X[\frac{m}{2}].
\end{align}
\end{example}
Then, we define a linear function over real field as follows:
\begin{align}
    L(X) = \sum^{\frac{m}{2}}_{j=1}X[j] 
\end{align}
with a bias term $B = - \frac{m}{2} + \frac{1}{2}$, where $L(X) + B = \frac{1}{2}$ if and only if $f(X) = 1$. Otherwise, $L(X) + B \leq - \frac{1}{2}$. Thus, we can compute $f(X)$ by computing its corresponding linear threshold function $\text{sgn} (L(X)+B)$, i.e., $f(X) = 1$ if $\text{sgn} (L(X)+B)=1$; otherwise, $f(X) = 0$ if $\text{sgn} (L(X)+B)=-1$. Unlike computing the function $f(X)$ with the degree $\frac{m}{2}$ which results in low security threshold, computing the linear function $L(X)$ allows us to apply a linear code on the computations which can lead to a much higher security threshold.
\subsection{Formal Description of Coded ANF}
Given the ANF representation defined in (\ref{eq:ANF}), we now present the proposed coded ANF scheme in the following. For each monomial $\prod_{j \in \mathcal{S}} X[j]$ such that $\mu_f(\mathcal{S}) = 1$, we define a linear function $L_{\mathcal{S}}: \mathbb{R}^m \rightarrow \mathbb{R}$ and a bias term $B_{\mathcal{S}}\in \mathbb{R}$ as follows:\footnote{The linear threshold function defined in (\ref{eq:LTF_ANF}) is adapted from the degree-$1$ polynomial threshold function $p(X) = \sum^m_{j=1}Z[j]X[j] -m +\frac{1}{2}$ considered in [17] where $X \in \{-1,1\}^m$ and $p(X)>0$ iff $X=Z$. Since the Boolean domain considered in~\cite{o2008extremal} is $\{-1,1\}$ instead of $\{0,1\}$ and all the bits are taken into account in $p(X)$, we define (\ref{eq:LTF_ANF}) by letting $Z[j] = 0, \forall j \notin \mathcal{S}$ and the bias term to be $-|\mathcal{S}|+\frac{1}{2}$ such that only the bits $X[j], \forall j \in \mathcal{S}$ in the domain $\{0,1\}$ are taken into account in (\ref{eq:LTF_ANF}). }
\begin{align}
    L_{\mathcal{S}}(X) =  \sum_{j \in \mathcal{S}}X[j], \quad B_{\mathcal{S}} = -|\mathcal{S}| + \frac{1}{2}. \label{eq:LTF_ANF}
\end{align}
It is clear that $L_{\mathcal{S}}(X)+B_{\mathcal{S}}=\frac{1}{2}$ if and only if $\prod_{j \in \mathcal{S}} X[j]=1$. Otherwise, $L_{\mathcal{S}}(X) + B_{\mathcal{S}}\leq -\frac{1}{2}$. Thus, there are $r(f)$ constructed linear threshold functions, and each monomial $\prod_{j \in \mathcal{S}} X[j]$ can be computed by its corresponding linear threshold function $\text{sgn} (L_{\mathcal{S}}(X)+B_{\mathcal{S}})$. 

By considering each bit in real field, the master encodes $X_1,X_2,\dots,X_K$ to $\tilde{X}_1,\tilde{X}_2,\dots,\tilde{X}_N$ using an $(N,K)$ MDS code. Each worker $n \in [N]$ stores $\tilde{X}_{n}$ locally. Each worker $n \in [N]$ computes the functions $\{L_{\mathcal{S}}(\tilde{X}_n)\}_{\{\mathcal{S} \subseteq [m],\mu_f(\mathcal{S})=1\}}$ and then sends the results back to the master. After receiving the results from the workers, the master first recovers $L_\mathcal{S}(X_k)$ for each $k \in [K]$ and each $\mathcal{S} \in \{\mathcal{G} : \mathcal{G} \subseteq [m], \mu_f(\mathcal{G}) = 1\}$. Then, the master has $\prod_{j \in \mathcal{S}} X_k[j] = 1$ if $\text{sgn} (L_{\mathcal{S}}(X_k)+B_{\mathcal{S}}) = 1$; $\prod_{j \in \mathcal{S}} X_k[j] = 0$ if $\text{sgn} (L_{\mathcal{S}}(X_k)+B_{\mathcal{S}}) = -1$. Lastly, the master recovers $f(X_1),\dots,f(X_K)$ by summing the monomials. Since each of $r(f)$ linear functions has up to $m$ variables, the complexity at each worker is $O(mr(f))$.
\begin{remark}
    We can demonstrate the decodability of $\{L_{\mathcal{S}}(\tilde{X}_n)\}_{n \in [N]}$'s by converting our problem to the distributed matrix-matrix multiplications as follows. Computing $\{L_{\mathcal{S}}(X_k)\}_{k \in [K]}$ for each $\mathcal{S}$ is equivalent to computing $K$ matrix-matrix multiplications $X_1A,X_2A,\dots,X_KA$ ($X_1,\dots, X_K$ are considered as row vectors) where $A$ is an $m$ by $|\mathcal{S}|$ matrix and each column of matrix $A$ is the coefficients of $X[j]$'s in the corresponding $L_{\mathcal{S}}(X)$. Similarly, computing $\{L_{\mathcal{S}}(\tilde{X}_n)\}_{n \in [N]}$ for the corresponding $\mathcal{S}$ is equivalent to computing $N$ matrix-matrix multiplications $\tilde{X}_1A,\tilde{X}_2A,\dots,\tilde{X}_NA$.         Therefore, our problem can be converted to the coded distributed matrix-matrix multiplication in which an $(N,K)$ MDS code is used to each element of the matrices $X_1,\dots,X_K$ and the encoded matrices $\tilde{X}_1,\dots,\tilde{X}_N$ are obtained. In~\cite{lee2018speeding}, it is shown that matrix multiplications $X_1A,X_2A,\dots, X_KA$ can be recovered from any $K$ out of $N$ coded results $\tilde{X}_1A,\dots \tilde{X}_NA$ by the MDS property and the linear property of matrix-matrix multiplications. In our problem, we deal with adversarial workers which are treated as errors. Since the system can be robust to $N-K$ erasures, one can show that the system can be robust to $\lfloor\frac{N-K}{2}\rfloor$ errors (adversaries) by Lemma $3$ proved in~\cite{yu2020straggler}. 
    \end{remark}
\subsection{Security Threshold of Coded ANF}
To decode the $(N,K)$ MDS code, coded ANF applies Reed-Solomon decoding. Successful decoding requires the number of errors of computation results such that $N \geq K+2b$. The following theorem shows that the security threshold provided by coded ANF is $\bigl \lfloor \frac{N-K}{2} \bigr \rfloor$ which is independent of $\textrm{deg}f$.
\begin{theorem} \label{thm:LTF_1}
Given a number of workers $N$ and a dataset $X = (X_1,\dots, X_K)$, the proposed coded ANF can be robust to $b$ adversaries for computing $\{f(X_k)\}^K_{k=1}$ for any Boolean function $f$, as long as
\begin{align}
    N\geq K + 2b;
\end{align}
i.e., coded ANF achieves the security threshold
\begin{align}
    \beta_{\textrm{ANF}} = \bigl \lfloor \frac{N-K}{2} \bigr \rfloor.
\end{align}
\end{theorem}
Whenever the master receives $N$ results from the workers, the master decodes the computation results using a length-$N$ Reed-Solomon code for each of $r(f)$ linear functions which incurs the total complexity $O(r(f)N \log^2{N} \log{\log{N}})$. Computing all the monomials via the signs of corresponding linear threshold functions incurs the complexity $O(Nr(f))$. Lastly, computing $f(X_1),\dots,f(X_K)$ by summing the monomials incurs the complexity $O(Nr(f))$ since there are $r(f)-1$ additions in function $f$. Thus, the total complexity of decoding step is $O(r(f)N \log^2{N} \log{\log{N}})$ which works well for small $r(f)$. Note that the operation of this scheme is over real field whose size does not scale with size of $m$.

%% file: 5-DNF.tex
In this section, we propose a coding scheme called \emph{coded Disjunctive normal form (DNF)} which computes the DNF representations of Boolean function by LTFs and a simple linear code is used for the data encoding. We start with an example to illustrate the idea behind coded DNF.
\begin{example}\label{ex:DNF}
Consider a function which has an ANF representation defined as follows:
\begin{align}
   f(X) = (X[1]\cdots X[m])\oplus (X[1] \oplus 1) \cdots (X[m] \oplus 1) \nonumber
\end{align}
which has the degree $\textrm{deg}f = m-1$ and the number of monomials $r(f) = 2^m-1$.
Alternatively, this function has a DNF representation as follows:
\begin{align}
    f(X) = (X[1] \wedge \cdots \wedge X[m]) \vee (\sim X[1]\wedge \cdots \wedge \sim X[m]) \nonumber
\end{align}
which has the weight $w(f) = 2$.
\end{example}
For the clause $X[1] \wedge \cdots \wedge X[m]$, we define a linear function over real field as follow:
\begin{align}
    L_1(X) = X[1]+\cdots+X[m]
\end{align}
with a bias term $B_1 = - m + \frac{1}{2}$, where $X[1] \wedge \cdots \wedge X[m] = 1$ if and only if $L_1(X) +B_1 = \frac{1}{2}$. Otherwise, $L_1(X)+B_1\leq - \frac{1}{2}$. Similarly, for the clause $ \sim X[1] \wedge \cdots \wedge \sim X[m]$, we define a linear function over real field as follows:
\begin{align}
    L_2(X) = -X[1]-\cdots-X[m]
\end{align}
with a bias $B_2 = \frac{1}{2}$, where $\sim X[1] \wedge \cdots \wedge \sim X[m] = 1$ if and only if $L_2(X) + B_2= \frac{1}{2}$. Otherwise, $L_2(X)+B_2\leq - \frac{1}{2}$.
Therefore, we can compute $f(X)$ by computing $\text{sgn} (L_1(X)+B_1)$ and $\text{sgn} (L_2(X)+B_2)$, i.e., $f(X) = 1$ if at least one of $\text{sgn} (L_1(X)+B_1)$ and $\text{sgn} (L_2(X)+B_2)$ is equal to $1$. Otherwise, $f(X) =0$. Unlike directly computing the function $f(X)$ with the degree of $m-1$, computing the linear functions $L_1(X)$ and $L_2(X)$ allows us to apply a linear code on the computations.
\subsection{Formal Description of Coded DNF}
Given the DNF representation defined in (\ref{eq:DNF}), we now present the proposed coded DNF scheme in the following. For each clause $T_i$ with the corresponding input $Y_i \in \textrm{Supp}(f)$ such that $f(Y_i)=1$, we define a linear function $L_i:\mathbb{R}^m \rightarrow \mathbb{R}$ and a bias term $B_i \in \mathbb{R}$ as follows:\footnote{Similar to the linear threshold function defined in (\ref{eq:LTF_ANF}), we define (\ref{eq:LTF_DNF}) by adjusting the bias term such that the threshold function can work in the domain of $\{0,1\}$. }
\begin{align}
    L_i(X) = \sum^m_{j=1}Z_i[j]X[j],\quad B_i = - \sum^m_{j=1}Y_i[j] + \frac{1}{2}  \label{eq:LTF_DNF}
\end{align}
where 
\begin{align}
Z_i[j] = 
    \begin{cases}
    1, \ & \text{if} \ Y_i[j]=1\\
    -1, \ &\text{if} \ Y_i[j]=0.
    \end{cases}
\end{align}
It is clear that $L_i(Y_i) + B_i= \frac{1}{2}$ and $L_i(X) +B_i\leq -\frac{1}{2}$ for all other inputs $X \neq Y_i$. Thus, there are $w(f)$ constructed linear threshold functions, and each clause $T_i$ can be computed by its corresponding linear threshold function $\text{sgn}(L_i(X)+B_i)$. 

By considering each bit over real field, the master encodes $X_1,X_2,\dots,X_K$ to $\tilde{X}_1,\tilde{X}_2,\dots,\tilde{X}_N$ using an $(N,K)$ MDS code. Each worker $n \in [N]$ stores $\tilde{X}_{n}$ locally. Each worker $n$ computes the functions $L_1(\tilde{X}_n),\dots,L_{w(f)}(\tilde{X}_n)$ and then sends the results back to the master. After receiving the results from the workers, the master first recovers $L_i(X_k)$ for each $i \in [w(f)]$ and each $k \in [K]$ via MDS decoding. Then, the master has $T_i(X_k) = 1$ if $\text{sgn} (L_i(X_k)+B_i) = 1$; otherwise $T_i(X_k) = 0$. Lastly, the master has $f(X_k)=1$ if at least one of $T_1(X_k),\dots,T_{w(f)}(X_k)$ is equal to $1$. Otherwise, $f(X_k)=0$. Since each of $w(f)$ linear functions has $m$ variables, the complexity at each worker is $O(mw(f))$.
\subsection{Security Threshold of Coded DNF}
Similar to coded ANF deploying Reed-Solomon code for the decoding process, we have the following theorem to show that the security threshold provided by coded DNF is $\bigl \lfloor \frac{N-K}{2} \bigr \rfloor$ which is independent of $\textrm{deg}f$.
\begin{theorem} \label{thm:LTF_2}
Given a number of workers $N$ and a dataset $X = (X_1,\dots, X_K)$, the proposed coded DNF can be robust to $b$ adversaries for computing $\{f(X_k)\}^K_{k=1}$ for any Boolean function $f$, as long as
\begin{align}
    N\geq K + 2b;
\end{align}
i.e., coded DNF achieves the security threshold
\begin{align}
    \beta_{\textrm{DNF}} = \bigl\lfloor \frac{N-K}{2} \bigr\rfloor.
\end{align}
\end{theorem}
Upon receiving $N$ results from the workers, the master decodes the computation results using a length-$N$ Reed-Solomon code for each of $w(f)$ linear functions which incurs the total complexity $O(w(f)N \log^2{N} \log{\log{N}})$. Computing all the clauses via the signs of corresponding linear threshold functions incurs the complexity $O(Nw(f))$.  Lastly, computing $f(X_1),\dots,f(X_K)$ by checking all the clauses requires the complexity $O(Nw(f))$. Thus, the total complexity of decoding step is $O(w(f)N \log^2{N} \log{\log{N}})$ which works well for small $w(f)$.
\begin{remark}
Learning the DNF representation of a Boolean function is an intensively studied problem in computational learning theory and is hard in general \cite{klivans2004learning}. Thus, people focus on some more tractable classes of functions, e.g., $O(\log{n})$-term DNF is considered in PAC learning literature \cite{kushilevitz1997simple}, which well motivates our proposed coded DNF.
\end{remark}
\begin{remark}
    Although both coded ANF and coded DNF achieve the security threshold $\bigl\lfloor \frac{N-K}{2} \bigr\rfloor$, coded ANF has the decoding complexity $O(r(f)N\log^2{N}\log{\log{N}})$ and coded DNF has the decoding complexity $O(w(f)N \log^2{N} \log{\log{N}})$. 
    Based on the sparsity $r(f)$ and the weight $w(f)$, one can choose either one of two schemes that has a smaller decoding complexity. When $r(f)$ is smaller than $w(f)$, coded ANF should be chosen. One the contrary, we can choose coded DNF. 
\end{remark}

%% file: 6-PTF.tex
In this section, we propose a coding scheme called \emph{coded polynomial threshold function (PTF)} which computes the DNF representations of Boolean function by PTFs and LCC is used for the data encoding. 
\subsection{Formal Description of Coded PTF}\label{subsec:construction}
Given the DNF representation defined in (\ref{eq:DNF}), we now present coded PTF. Following the construction proposed in \cite{o2008extremal,klivans2004learning}, we now construct a polynomial threshold function $\text{sgn}(P(X))$ for computing $f(X)$ where $P:\mathbb{R}^m \rightarrow \mathbb{R}$ is a polynomial function with the degree at most $\lfloor \log_2{w(f)} \rfloor+1$. The construction of such PTF has the following steps.
\begin{enumerate}[leftmargin=*]
    \item \textbf{Decision Tree Construction:} We construct an $w(f)$-leaf decision tree over variables $X[1],\dots,X[m]$ such that each input in $\text{Supp}(f)$ arrives at a different leaf. Such a tree can be always constructed by a greedy algorithm. Let $\ell_i$ be a leaf of this tree in which $Y_i$ reaches leaf $\ell_i$. We label $\ell_i$ with the linear threshold function $\text{sgn}(L_i(X)+B_i)$ where $L_i(X)$ and $B_i$ are defined in (\ref{eq:LTF_DNF}). The constructed decision tree, in which internal nodes are labeled with variables and leaves are labeled with linear threshold functions, computes exactly $f$.
    \item \textbf{Decision List Construction:} For this $w(f)$-leaf decision tree, we construct an equivalent $\lfloor \log_2{w(f)} \rfloor$-decision list. Following from the definition that the rank of an $w(f)$-leaf tree is at most $\lfloor \log_2{w(f)} \rfloor$. We find a leaf in the decision tree at distance at most $\lfloor \log_2{w(f)} \rfloor$ from the root, and place the literals along the path to the leaf as a monomial at the top of a new decision list. We then remove the leaf from the tree, creating a new decision tree with one fewer leaf, and repeat this process \cite{blum1992rank}. Without loss of generality, we let $\ell_i$ be the $i$-th removed leaf in the process of list construction with the corresponding monomial $C_i$ of at most $\lfloor \log_2{w(f)} \rfloor$ variables. The constructed list is defined as "if $C_1(X)=1$ then output $\frac{1+\text{sgn}(L_1(X)+B_1)}{2}$; else if $C_2(X)=1$ then output $\frac{1+\text{sgn}(L_2(X)+B_2)}{2}$; ... else if $C_{w(f)}(X)=1$ then output $\frac{1+\text{sgn}(L_{w(f)}(X)+B_{w(f)})}{2}$.
    \item \textbf{Polynomial Threshold Function Construction:} Having the constructed decision list, we now construct the polynomial function $P(X)$ with degree of at most $\lfloor \log{w(f)} \rfloor+1$ as follows:
    \begin{align}
       P(X) &= A_1C_1(X)(L_1(X)+B_1) +  \dots \nonumber \\ 
       &+ A_{w(f)}C_{w(f)}(X)(L_{w(f)}(X)+B_{w(f)}) \nonumber
    \end{align}
 where $A_1 \gg A_2 \gg A_3  \dots \gg A_m > 0$ are appropriately chosen positive values.
\end{enumerate}

After constructing the corresponding PTF $\text{sgn}(P(X))$ for Boolean function $f(X)$, the procedure of computations is as follows. By considering each bit over real field, the master encodes $X_1,X_2,\dots,X_K$ to $\tilde{X}_1,\tilde{X}_2,\dots,\tilde{X}_N$ using LCC. Each worker $n \in [N]$ stores $\tilde{X}_n$ locally. Each worker $n$ computes the function $P(\tilde{X}_n)$ and then sends the result back to the master. After receiving the results from the workers, the master first recovers $P(X_1),\dots,P(X_K)$ via LCC decoding. Then, the master has $f(X_k) = 1$ if $\text{sgn} (P(X_k)) = 1$; otherwise $f(X_k) = 0$. Since $C_i(X)$'s are monomials with the degree of at most $\lfloor \log_2{w(f)} \rfloor$, computing $A_iC_i(X)$ incurs the complexity $O(\lfloor \log_2{w(f)} \rfloor)$. Also, computing $L_i(X)+B_i$ incurs the complexity $O(m)$. Thus, computing function $P(X)$ at each worker incurs the total complexity $O(w(f)(\lfloor\log_2{w(f)}\rfloor+m))$.  
\subsection{Security Threshold of Coded PTF}
Since $P(X)$ has degree of at most $\lfloor \log_2{w(f)} \rfloor+1$, to be robust to $b$ adversaries, LCC requires the number of workers $N$ such that $N\geq (K-1)(\lfloor \log_2{w(f)} \rfloor+1) +2b+1$. Then, we present the security threshold provided by coded PTF in the following theorem.
\begin{theorem} \label{thm:PTF}
Given a number of workers $N$ and a dataset $X = (X_1,\dots, X_K)$, the proposed coded polynomial threshold function can be robust to $b$ adversaries for computing $\{f(X_k)\}^K_{k=1}$ for any Boolean function $f$, as long as
\begin{align}
    N\geq (K-1)(\lfloor \log_2{w(f)} \rfloor+1)+2b+1;
\end{align}
i.e., coded PTF achieves the security threshold
\begin{align}
   \beta_{\textrm{PTF}} = \bigl\lfloor\frac{N - (K-1)(\lfloor \log_2{w(f)} \rfloor+1)-1}{2}\bigr\rfloor.
\end{align}
\end{theorem}
Whenever the master receives $N$ results from the workers, the master decodes the computation results using a length-$N$ Reed-Solomon code for the polynomial function which incurs the total complexity $O(N \log^2{N} \log{\log{N}})$. Lastly, computing $f(X_1), f(X_2),\dots,f(X_K)$ by checking the signs requires the complexity $O(N)$. Thus, the total complexity of decoding step is $O(N \log^2{N} \log{\log{N}})$. 

In the following example, we show that coded PTF outperforms LCC for the Boolean functions with the polynomial size of $r(f)$ and $w(f)$.
\begin{example}\label{ex:PTF}
Consider a function which has an ANF representation defined as follows:
\begin{align}
   f(X) & = (X[1]\oplus X[2])\cdots(X[2m'-1]) \oplus X[2m']) \nonumber\\ &\times X[2m'+1]\cdots X[m]
\end{align}
where $m' = \lfloor \log_2{m^2} \rfloor$. Note that here we focus on the case that $m$ is large enough such that $m>m' = \lfloor \log_2{m^2} \rfloor$. The function $f$ has the degree of $m-\lfloor \log_2{m^2} \rfloor$, the sparsity of $\approx m^2$ and the weight of $\approx m^2$.
\end{example}
For the Boolean function considered in Example \ref{ex:PTF}, coded PTF achieves the security threshold $\bigl \lfloor \frac{N - (K-1)(\lfloor \log_2{m^2} \rfloor+1) -1}{2} \bigr \rfloor$ which is greater than the security threshold $\bigl \lfloor \frac{N - (K-1)(m-\lfloor \log_2{m^2} \rfloor) -1}{2} \bigr \rfloor$ provided by LCC. Although coded ANF and coded DNF achieve security threshold $\bigl \lfloor\frac{N-K}{2}\bigr \rfloor$ but they require decoding complexity $O(m^2N \log^2{N} \log{\log{N}})$ which has the order of $m^2$, i.e., they only work for small $m$. With the security slightly worse than coded ANF and coded DNF, coded PTF achieves the better decoding complexity which is independent of $m$, i.e., coded PTF can work for large $m$.

\subsection{Coded $D$-partitioned PTF}
In this subsection, we extend coded PTF by proposing \emph{coded $D$-partitioned polynomial threshold function} whose idea is to partition the Boolean function into some DNFs and construct their corresponding PTFs with low-degree. It allows us to apply LCC on the corresponding low-degree PTFs for improving the security threshold. 

Given the DNF representation defined in (\ref{eq:DNF}) of Boolean function $f$ and an integer $D$ ($1\leq D \leq w(f)$), we partition the DNF representation of $f$ to $D$ different DNF representations as follows:
\begin{align}
    f = \mathcal{G}_1 \vee \mathcal{G}_2 \vee \dots \vee \mathcal{G}_D 
\end{align}
where each $\mathcal{G}_d$ includes $\frac{w(f)}{D}$ clauses of $m$ literals, e.g.,
\begin{align}
    \mathcal{G}_1 = T_1 \vee \dots \vee T_{\frac{w(f)}{D}}.
\end{align}
Thus, we have that each $\mathcal{G}_d$ is a Boolean function with weight of $\frac{w(f)}{D}$. By the PTF construction described in Subsection \ref{subsec:construction}, each Boolean function $\mathcal{G}_d$ can be computed by a PTF $\text{sgn}(P_d(X))$ where $P_d(X)$ has degree of at most $\lfloor \log_2{\frac{w(f)}{D}} \rfloor +1$.

Similar to coded PTF using LCC for data encoding, each worker $n \in [N]$ stores $\tilde{X}_n$ locally. Each worker $n$ computes the function $P_1(\tilde{X}_n),\dots,P_D(\tilde{X}_n)$ and then sends the results back to the master. Upon receiving the results from the workers, the master first recovers $P_d(X_1),\dots,P_d(X_K)$ for each $d$ via LCC decoding. Then, the master has $f(X_k) = 1$ if at least one of $\text{sgn} (P_1(X_k)),\dots,\text{sgn}(P_D(X_k))$ is equal to $1$. Otherwise, $f(X_k) = 0$. Similar to coded PTF, computing $D$ polynomial functions with the degree up to $\lfloor \log_2{\frac{w(f)}{D}} \rfloor+1$ at each worker incurs the complexity $O(w(f)(\lfloor\log_2{\frac{w(f)}{D}}\rfloor+m))$. 

Since each $P_d(X)$ has degree of at most $\lfloor \log_2{\frac{w(f)}{D}} \rfloor+1$, to be robust to $b$ adversaries, LCC requires the number of workers $N$ such that $N\geq (K-1)(\lfloor \log_2{\frac{w(f)}{D}} \rfloor+1) +2b+1$. Formally, we have the following theorem.
\begin{theorem} \label{thm:PTF_D}
Given a number of workers $N$ and a dataset $X = (X_1,\dots, X_K)$, the proposed coded $D$-partitioned polynomial threshold function can be robust to $b$ adversaries for computing $\{f(X_k)\}^K_{k=1}$ for any Boolean function $f$, as long as
\begin{align}
    N\geq (K-1)(\lfloor \log_2{\frac{w(f)}{D}} \rfloor+1)+2b+1;
\end{align}
i.e., coded $D$-partitioned PTF achieves the security threshold
\begin{align}
   \beta_{\textrm{PTF}}(D) = \bigl\lfloor \frac{N - (K-1)(\lfloor \log_2{\frac{w(f)}{D}} \rfloor+1)-1}{2} \bigr\rfloor.
\end{align}
\end{theorem}
Whenever the master receives $N$ results from the workers, the master decodes the computation results using a length-$N$ Reed-Solomon code for $D$ constructed polynomial function which incurs the total complexity $O(DN \log^2{N} \log{\log{N}})$. Then, computing $f(X_1), f(X_2),\dots,f(X_K)$ by checking the signs and OR operations requires the complexity $O(DN)$. Thus, the total complexity of decoding step is $O(DN \log^2{N} \log{\log{N}})$.
\begin{remark}
The proposed coded $D$-partitioned PTF characterize a tradeoff between the security threshold and the decoding complexity. For each chosen $D (1\leq D \leq w(f))$, the pair of the security threshold and the decoding complexity $(\bigl\lfloor \frac{N - (K-1)(\lfloor \log_2{\frac{w(f)}{D}} \rfloor+1)-1}{2} \bigr\rfloor, DN \log^2{N} \log{\log{N}})$ can be achieved by the proposed coded $D$-partitioned PTF. In particular, the proposed coded DNF and coded PTF schemes correspond to the two extreme points of this tradeoff that minimize the security threshold and the decoding complexity respectively. Coded DNF corresponds to the point $D=1$, i.e., no partition performed. On the other hand, coded corresponds to the point $D = w(f)$, i.e., each DNF after partition process only contains one vector in $\{0,1\}^m$. Thus, coded $D$-partitioned PTF generalizes our previously proposed coded DNF and coded PTF, and allows to systematically operate at any points on this tradeoff.
\end{remark}

\begin{remark}
The total complexity of computing $K$ evaluations $f(X_1),\dots,f(X_K)$ via ANF is $O(Kr(f) \textrm{deg} f)$. Thus, it is more efficient to use coded ANF than computing all the evaluations at the master when $\textrm{deg} f > \frac{N}{K}\log^2{N} \log{\log{N}}$. On the other hand, since computing $f(X_1),\dots,f(X_K)$ via DNF incurs the total complexity $O(Kmw(f))$, we can conclude that it is more efficient to use coded DNF when $m > \frac{N}{K}\log^2{N} \log{\log{N}}$. When $mw(f) > \frac{DN}{K}\log^2{N} \log{\log{N}}$, coded $D$-partitioned PTF is more efficient than computing all the evaluations at the master.
    \end{remark}

%% file: 7-lowerbound.tex
In this section, we show that coded ANF and coded DNF are optimal in terms of the security threshold. We start by defining the recovery threshold and the hamming distance of a scheme as follows:
\begin{definition}
For any integer $k$, we say a scheme is $k$-recoverable if the master can recover $h(X_1),\dots,h(X_K)$ given the computing results from any $k$ workers. We define the recovery threshold of a scheme $(\vec{g},h)$, denoted by $K(\vec{g},h)$, as the minimum integer $k$ such that scheme $(\vec{g}, h)$ is $k$-recoverable.
\end{definition}
\begin{definition}
We define the Hamming distance of any scheme $(\vec{g},h)$, denoted by $d(\vec{g},h)$, as the maximum integer $d$ such that for any pair of input dataset whose computation results $h(X_1),\dots,h(X_K)$ are different, at least $d$ workers compute different values of $h(\tilde{X}_n)$.
\end{definition}
We prove the matching outer bound for coded ANF and coded DNF by the following theorem whose proof can be found in Appendix A. 
\begin{theorem}\label{thm:lowerbound}
For a distributed computing problem of computing Boolean function $f$ using $N$ workers over a dataset $X = (X_1,\dots, X_K)$, any scheme $(\vec{g},h)$ can achieve the security threshold up to
\begin{align}
    \beta^{*} = \bigl\lfloor \frac{N-K}{2} \bigr\rfloor.
\end{align}
\end{theorem}

By Theorem \ref{thm:lowerbound}, we have shown that the proposed coded ANF and coded DNF schemes are optimal in terms of the security threshold.


%% file: 8-example.tex
\begin{table*}[t]
  \centering
  \scalebox{0.85}{
  \begin{tabular}{| c | c | c | c | c | c | c | c | c | c | c | c| c| c| c| c| c|}
    \hline
    \rule{0pt}{12pt} $X$ & $0000$ & $0001$ & $0010$ & $0011$ & $0100$ & $0101$ & $0110$ & $0111$ & $1000$ & $1001$ & $1010$ & $1011$ & $1100$ & $1101$ & $1110$ & $1111$\\ \hline
    \rule{0pt}{12pt} $s(X)[1]$ & $1$ & $0$ & $1$ & $0$ & $0$ & $1$ & $1$ & $0$ &$0$ &$1$ &$1$ &$0$ &$0$ &$0$ &$1$ &$1$ \\ \hline
    \rule{0pt}{12pt} $s(X)[2]$ & $1$ & $1$ & $1$ & $0$ & $1$ & $0$ & $1$ & $0$ &$0$ &$1$ &$0$ &$1$ &$0$ &$1$ &$0$ &$0$ \\ \hline
    \rule{0pt}{12pt} $s(X)[3]$ & $1$ & $1$ & $0$ & $1$ & $1$ & $1$ & $1$ & $0$ &$0$ &$0$ &$0$ &$0$ &$1$ &$0$ &$0$ &$1$ \\ \hline
    \rule{0pt}{12pt} $s(X)[4]$ & $1$ & $1$ & $1$ & $1$ & $0$ & $1$ & $0$ & $1$ &$0$ &$0$ &$1$ &$1$ &$0$ &$0$ &$0$ &$0$ \\ \hline
  \end{tabular}
  }
  \caption{An example of a $4$-bit S-box. Each coordinate of $s(X)$ can be represented by a degree-$3$ ANF representation. }\label{table:s_box}
\end{table*}
To demonstrate the impact of the proposed schemes, we consider a cryptosystem which is designed to enable two parties to securely communicate over an insecure channel~\cite{carlet2010boolean}. In a cryptosystem, the plaintext is encrypted to the cyphertext before the communication from one user to the another user, e.g., one user of a party shares the same secret key with the user of another party to communicate secretly. Since the security of symmetric cryptosystems is strongly influenced by Boolean functions, many properties of Boolean functions must be utilized (e.g., high nonlinearity, high algebraic degree, and etc) in order to resist the known mathematical attacks. More specifically, a cipher must not be well-approximated by linear functions to be secure against linear attacks~\cite{matsui1993linear}. High algebraic degree of Boolean function increases the linear complexity in block ciphers and result in more complicated systems of equations describing the cipher which make structural attacks of the cipher more difficult~\cite{Bajric19}. 

In particular, we focus on one of subclasses of symmetric key cryptosystem: block cyphers. As the non-linear component in most block ciphers, \emph{S-boxes} are one of the most important building blocks in symmetric cryptography and chosen to be cryptographically strong enough against the attacks. Formally, an S-box $s:\{0,1\}^m \rightarrow \{0,1\}^m$ is represented by a collection of $m$ Boolean functions of $m$ input bits, and each Boolean function is one of the coordinates of function $s$. Please see Table~\ref{table:s_box} for an example of a $4$-bit S-box.

Each coordinate of $s(X)$ presented in the example in Table~\ref{table:s_box} can be represented by a degree-$3$ ANF representation as follows:
 \begin{align}
     s(&X)_1  =  1  \oplus X[1] \oplus X[3] \oplus X[4] \oplus X[2]X[3]  \oplus X[2]X[4] \nonumber\\
     &\oplus X[3]X[4] \oplus X[1]X[3]X[4]  \oplus X[2]X[3]X[4]\\
     s(&X)_2 =  1 \oplus X[4]\oplus X[1]X[2] \oplus X[1]X[3]  \oplus X[1]X[4] \nonumber\\ &\oplus X[1]X[2]X[3] \oplus X[1]X[2]X[4]
     \oplus X[1]X[3]X[4]\\
     s(&X)_3 =  1 \oplus X[2]  \oplus X[4]\oplus X[1]X[2] \oplus X[2]X[3] \nonumber \\ 
     &\oplus X[3]X[4] \oplus X[2]X[4] \oplus X[1]X[2]X[4] \nonumber\\ & \oplus X[1]X[3]X[4]\\
     s(&X)_4 =  1 \oplus X[3] \oplus X[4] \oplus X[1]X[3]\oplus X[2]X[4] \nonumber\\ &\oplus X[3]X[4] \oplus X[1]X[3]X[4] \oplus X[2]X[3]X[4].
 \end{align}

Desirably, S-box functions are designed such that the degree of the polynomial in S-box is large, which makes more difficult the application of higher order differential attacks. As the size of datasets grows, it is necessary to take advantage of the power of distributed computing, i.e., the data encryption computations are computed in a distributed manner. Let us consider the encryption problem of computing a $8$-bit S-box function over a dataset $X=(X_1,\dots,X_{10})$ using a system of $N=100$ workers. The best possible degree of $8$-bit S-box $\textrm{deg}s$ is equal to $7$~\cite{boss2016strong}. For computing $s(X_1),\dots,s(X_{10})$ distributedly, the security threshold achieved by LCC is $\bigl \lfloor \frac{N-(K-1)\textrm{deg}s-1}{2}\bigr \rfloor = 18$. Our proposed coded ANF and coded DNF provide the optimal security threshold of $\bigl \lfloor \frac{N-K}{2} \bigr \rfloor = 45$. As compared to LCC, the proposed coded ANF and coded DNF schemes improve the security threshold by $150\%$.

%% file: 9-sparse_poly.tex
In this section, we extend our problem to a more general computation model. More specifically, we focus on computing multivariate polynomial $f: \mathbb{V} \rightarrow \mathbb{U}$ over a dataset $X_1,\dots,X_K$ using  a master and $N$ workers, where $\mathbb{V}$ and $\mathbb{U}$ are arbitrary vector spaces over the certain field $\mathbb{F}$. We denote by $r(f)$ the number of monomials appearing in $f(X)$\footnote{Similar to the case of Boolean functions, the total complexity of computing $f(X_1),\dots,f(X_K)$ is $O(Kr(f)\textrm{deg}f )$.}.

As we see in the problem of computing Boolean functions, the security threshold provided by LCC can be low if $\textrm{deg} f$ is high. To resolve such high degree difficulty which arises in computing general polynomials, we propose two different schemes: \textit{coded data logarithm} and \textit{coded data augmentation}. Especially, the proposed coded data logarithm scheme reduces the degree of polynomial computations by computing the logarithm of original data; and the proposed coded data augmentation reduces the degree of polynomial computations by pre-storing some low-degree monomials in advance.
\subsection{Coded Data Logarithm}
First, we illustrate the idea behind coded data logarithm by the following example.
\begin{example}
Consider the problem of computing function $f(X) = X^2$ in real field using $3$ workers over a dataset $\vec{X} = (X_1,X_2)$, where input $X_i$'s are $2 \times 2$ matrices. 

We start by constructing a degree-$1$ multivariate polynomial for the function $f(X)$. The function $f(X) = X^2$ can be explicitly written as follows:
\begin{align*}
    f(X) = 
    \begin{bmatrix}
[X]^2_{11} + [X]_{12}[X]_{21} & [X]_{11}[X]_{12} + [X]_{12}[X]_{22}\\
[X]_{11}[X]_{21} + [X]_{21}[X]_{22} & [X]_{12}[X]_{21} + [X]^2_{22}
\end{bmatrix}
\end{align*}
which includes $7$ monomials: 
\begin{align*}
    [X]^2_{11}, \quad [X]^2_{22}, \quad [X]_{12}[X]_{21}, \quad [X]_{11}[X]_{12}, \\ [X]_{12}[X]_{22}, \quad [X]_{11}[X]_{21}, \quad [X]_{21}[X]_{22}.
\end{align*}
By taking the logarithm of the absolute value of each monomial appearing in $f(X)$, we have
\begin{align*}
    2&\log{|[X]_{11}|}, \quad 2\log{|[X]_{22}|}, \quad  \log{|[X]_{12}|}+\log{|[X]_{21}|},\\
    &\log{|[X]_{11}|}+\log{|[X]_{12}|},\quad \log{|[X]_{12}|}+\log{|[X]_{22}|},\\ &\log{|[X]_{11}|}+\log{|[X]_{21}|},\quad  \log{|[X]_{21}|}+\log{|[X]_{22}|},
\end{align*}
which can be rewritten as:
\begin{align*}
        2[W]_{11}, \ 2[W]_{22}, \ [W]_{12}+[W]_{21}, \ [W]_{11}+[W]_{12},\\ [W]_{12}+[W]_{22},  [W]_{11}+[W]_{21}, \ [W]_{21}+[W]_{22},
\end{align*}
where $[W]_{ij} = \log{|[X]_{ij}|}$. We define a degree-$1$ multivariate polynomial $h(W)$ as follows: 
\begin{align*}
    h(W) = \Bigl [ 2[W]_{11}, \ 2[W]_{22}, \ [W]_{12}+[W]_{21}, \ [W]_{11}+[W]_{12},\\ [W]_{12}+[W]_{22}, \ [W]_{11}+[W]_{21}, \ [W]_{21}+[W]_{22} \Bigr ].
\end{align*}

To take advantage of the function $h(W)$ with the degree of $1$, we take the logarithm of each entry's absolute value in $X_1$ and $X_2$ and define two matrices $W_1$ and $W_2$ as follows:
\begin{align}
&W_1 = 
    \begin{bmatrix}
\log|[X_1]_{11}|& \log|[X_1]_{12}|\\
\log|[X_1]_{21}|& \log|[X_1]_{22}|
\end{bmatrix}, \quad  \nonumber \\ 
&W_2 = 
    \begin{bmatrix}
\log|[X_2]_{11}|& \log|[X_2]_{12}|\\
\log|[X_2]_{21}|& \log|[X_2]_{22}|
\end{bmatrix}.
\end{align}
Then, we encode $W_1$ and $W_2$ to $\tilde{W}_1$, $\tilde{W}_2$ and $\tilde{W}_3$ using an $(3,2)$ MDS code. Each worker $n$ computes $h(\tilde{W}_n)$ where each entry of $h(W)$ is a linear combination of the logarithm of the corresponding $X$'s entries' absolute values. By calculating the exponential of each entry in $h(W)$, the master can obtain the absolute values of all monomials appearing in $f(X)$, e.g., $[W]_{12} + [W]_{21} = \log |[X]_{12}|+\log |[X]_{21}| = \log |[X]_{12}[X]_{21}|$. 

Computing the degree-$1$ (linear) function $h(W)$ allows us to apply  a simple linear code to achieve the optimal security threshold.
\end{example}
In the following, we formally present the proposed coded data logarithm scheme. Given any multivariate polynomial function $f:\mathbb{V} \rightarrow \mathbb{U}$ over real field, the proposed coded data logarithm scheme first constructs the logarithmic data and a degree-$1$ multivariate polynomial function $h$ by the followings:
\begin{enumerate}[leftmargin=*]
    \item \textbf{Logarithmic Data Construction:} For each $X_k$, we construct a logarithmic data $W_k$ where each entry of $W_k$ is the logarithm of $X_k$ 's corresponding entry's absolute value\footnote{Note that if there is any entry of $X_1,\dots,X_K$ is zero, we can replace that entry by a non-zero value and proceed the proposed scheme. Since the monomials with a zero entry is always equal to zero, we can set them to zero in the decoding process.}, i.e., $W_k[j] = \log |X_k[j]|$ where we denote by $X_k[j]$ the $j$-th input value of $X_k$ without loss of generality.
    \item \textbf{Degree-$1$ Multivariate Polynomial Construction:} Construct a multivariate polynomial function $h(W)$ with degree of $1$ which computes the logarithm of absolute values of all monomials appearing in $f(X)$, i.e., for each monomial $\prod_{j \in \mathcal{S}}X[j]$ appearing in $f(X)$, the function $h(W)$ computes $\sum_{j \in \mathcal{S}}\log{|X[j]|} = \sum_{j \in \mathcal{S}}W[j]$.
\end{enumerate}
After the construction of corresponding logarithmic data $W_1,\dots,W_K$ for $X_1,\dots,X_K$, the procedure of computations is as follows. The master encodes $W_1,\dots,W_K$ to $\tilde{W}_1, \dots, \tilde{W}_N$ using an $(N,K)$ MDS code. Each worker $n$ computes $h(\tilde{W}_n)$ and then sends the result back to the master. Upon receiving all results from the workers, the master first recovers $h(W_1),\dots,h(W_k)$ and calculates the exponential of each entry of $h(W_k)$ which recovers the absolute values of all monomials appearing in $f(X_k)$. Then, each monomial term can be determined by changing the sign accordingly. Lastly, the master recovers $f(X_1),\dots,f(X_K)$ by summing the monomial terms and the bias terms. Since each of $r(f)$ monomials has the degree up to $\textrm{deg}f$, the complexity at each worker is $O((\textrm{deg}f) r(f))$. 

 Reed-Solomon decoding is used for decoding the $(N,K)$ MDS code. Successful decoding requires the number of errors of computation results such that $N \geq K+2b$. The following theorem shows the security threshold achieved by the proposed coded data logarithm scheme.
\begin{theorem} \label{thm:log}
Given a number of workers $N$ and a dataset $X = (X_1,\dots, X_K)$, the proposed coded data logarithm scheme can be robust to $b$ adversaries for computing $\{f(X_k)\}^K_{k=1}$ for any multivariate polynomial $f$, as long as
\begin{align}
    N\geq K + 2b;
\end{align}
i.e., coded data logarithm achieves the security threshold
\begin{align}
    \beta_{\textrm{LOG}} = \bigl \lfloor \frac{N-K}{2} \bigr \rfloor.
\end{align}
\end{theorem}
Using a length-$N$ Reed-Solomon code for each of $r(f)$ linear functions incurs the total complexity $O(r(f)N \log^2{N} \log{\log{N}})$. Computing the exponential of all the monomials incurs the complexity $O(Nr(f))$. Lastly, computing $f(X_1),\dots,f(X_K)$ by summing the monomials incurs the complexity $O(Nr(f))$. Thus, the total complexity of decoding step is $O(r(f)N \log^2{N} \log{\log{N}})$. Coded  data logarithm provides the optimal security threshold, and has low decoding complexity for computing the sparse polynomials (small $r(f)$).

\subsection{Coded Data Augmentation}
In the following example, we show how the proposed coded data augmentation scheme reduces the degree of polynomial computations.
\begin{example}
Consider the problem of computing a multivariate polynomial function $f$ with degree of $8$ defined as follows:
\begin{align}
    f(X) = x_1^5x_2^3 + x_2x_3^3 + 2
\end{align}
where each input $X$ has three entries $x_1,x_2,x_3$.

To reduce the degree of computation such that using LCC can be robust to more adversaries in the system, we augment each input $X$ by adding all degree-$2$ monomials as follows:
\begin{align}
    \bar{X}  & = [x_1 \ x_2 \ x_3 \ x^2_1 \ x^2_2 \ x^2_3 \ x_1x_2 \ x_1x_3 \ x_2x_3] \nonumber\\ 
    &= [x_1 \ x_2 \ x_3 \ y_1 \ y_2 \ y_3 \ y_4 \ y_5 \ y_6].
\end{align}
With data augmentation above, computing $f(X)$ is equivalent to computing $h(\bar{X})$ defined as follows:
\begin{align}
    h(\bar{X}) = y^2_1y_2y_4 + y_3y_6 +2 
\end{align}
which is the function with degree of $4$.

By prestoring the twice amount of data in each worker, the system can be robust to number of $\frac{4(K-1)}{2} = 2(k-1)$ more adversaries using LCC. Such pre-storing some low-degree polynomials enable us to enhance the robustness against Byzantine workers in the system.
\end{example}

In the following, we formally present the proposed coded data augmentation scheme. Given any multivariate polynomial $f:\mathbb{V} \rightarrow \mathbb{U}$ over a field $\mathbb{F}$ with an integer $q$, coded data augmentation first augments data and construct a low degree polynomial as follows:
\begin{enumerate}[leftmargin=*]
    \item \textbf{Data Augmentation:} For each $X_k$, we construct $\bar{X}_k$ by adding all the monomials of $X_k$'s entries with the degree up to $q$, i.e, adding $\prod_{j \in \mathcal{S}}X[j]$ for all $\mathcal{S} \subseteq [q]$.
    \item \textbf{Low Degree Polynomial Construction:} By substituting each added monomial as a new variable, we construct a multivariate polynomial function $h(\bar{X}_k)$ with degree of $u+\mathbbm{1}_{\{r>0\}}$, in which degree of $f$ can be uniquely written as $\textrm{deg} f = qu+r$ and $0 \leq r \leq q-1$. We note that such constructed polynomial is not unique but degree of $h$ is unique.
\end{enumerate}
The procedure of computations is as follows. The master encodes $\bar{X}_1,\dots,\bar{X}_K$ to $\tilde{X}_1,\dots,\tilde{X}_N$ using LCC encoder. Each worker $n$ computes $h(\tilde{X}_n)$ and then sends the result back to the master. Whenever the master receives $N$ results from the workers, the master recovers $h(\bar{X}_1),\dots,h(\bar{X}_K)$ using a length-$N$ Reed-Solomon code. Lastly, the master has $f(X_1) = h(\bar{X}_1), f(X_2) = h(\bar{X}_2),\dots,f(X_K) = h(\bar{X}_K)$. Since each of $r(f)$ monomials has the degree up to $u+\mathbbm{1}_{\{r>0\}}$, the complexity at each worker is $O((u+\mathbbm{1}_{\{r>0\}}) r(f))$. 

Because the constructed function $h(\bar{X})$ has degree of $u+\mathbbm{1}_{\{r>0\}}$ ($\textrm{deg} \ f = qu+r$), to be robust to $b$ adversaries, LCC requires the number of workers $N$ such that $N\geq (K-1)(u+\mathbbm{1}_{\{r>0\}}) +2b+1$. Then, we have the following theorem.
\begin{theorem} \label{thm:aug}
Given a number of workers $N$ and a dataset $X = (X_1,\dots, X_K)$, the proposed coded data augmentation scheme with parameter $q$ can be robust to $b$ adversaries for computing $\{f(X_k)\}^K_{k=1}$ for any multivariate polynomial $f$, as long as
\begin{align}
    N\geq (K-1)(u+\mathbbm{1}_{\{r>0\}}) + 2b +1;
\end{align}
i.e., coded data augmentation with parameter $q$ achieves the security threshold
\begin{align}
    \beta_{\textrm{AUG}} = \bigl \lfloor \frac{N-(K-1)(u+\mathbbm{1}_{\{r>0\}})-1}{2} \bigr \rfloor.
\end{align}
where $\textrm{deg} f = qu +r$ and $0 \leq r \leq q-1$.
\end{theorem}
Decoding the computation results using a length-$N$ Reed-Solomon code for the constructed polynomial function incurs the total complexity $O(N \log^2{N} \log{\log{N}})$. By trading the cost of storing more data for improving robustness against adversarial workers, coded data augmentation can be applied to any multivariate general polynomials and robust to $(K-1)(\textrm{deg}f-u-\mathbbm{1}_{\{r>0\}})/2$ more adversaries than LCC. 
    \begin{remark}
   Since computing all $K$ evaluations $f(X_1),\dots,f(X_K)$ at the master incurs the total complexity $O(Kr(f) \textrm{deg} f)$, it is more efficient to use coded data logarithm when $\textrm{deg} f > \frac{N}{K}\log^2{N} \log{\log{N}}$. When $r(f)\textrm{deg}f > \frac{N}{K}\log^2{N} \log{\log{N}}$, coded data augmentation is more efficient than computing all the evaluations at the master.
    \end{remark}

%% file: 10-conclusion.tex
In this paper, we focus on computing a Boolean function in a distributed manner against adversarial servers. To resolve the degree problem of using LCC (i.e., the security threshold provided by LCC can be low if the polynomial's degree is high), the proposed schemes called coded ANF, coded DNF and coded PTF largely improve the security threshold by modeling the polynomial as the concatenation of some low-degree polynomial functions and threshold functions. It is shown that coded ANF and coded DNF are optimal by matching to the derived theoretical outer bound; and increase the security threshold by $150\%$ for computing $8$-bit S-box in the application of block cyphers using a distributed computing system with $100$ workers.

There are many interesting directions can be pursued on the problem of coded Boolean computations. For example, the proposed coded ANF and coded DNF
require embedding bits to reals, which might lead to some floating-point errors during decoding process. Thus, one direction is to implement two schemes in an actual computing system and measure the effect of field transformation.

%% file: Appendix-proof.tex
The following lemma (Lemma 3 in [35]) is presented to bridge the coding theory and distributed computing via the recovery threshold and the hamming distance of a scheme. 
\begin{lemma}\label{lemma:hamming}
For any scheme $(\vec{g},h)$, we have
\begin{align}
    K(\vec{g},h) & = N-d(\vec{g},h) +1,\\
    E_{\text{detect}}(\vec{g},h) & = d(\vec{g},h)-1,   \\
    E_{\text{correct}}(\vec{g},h) & = \bigl \lfloor \frac{d(\vec{g},h)-1}{2} \bigr \rfloor
\end{align}
where $K(\vec{g},h)$ is the the recovery threshold provided by scheme $(\vec{g},h)$, $E_{\text{detect}}(\vec{g},h)$ denotes the maximum number of errors can be detected by the scheme, and $E_{\text{correct}}(\vec{g},h)$ denotes the maximum number of errors can be corrected by the scheme.
\end{lemma}
Lemma \ref{lemma:hamming} indicates that given any scheme that achieves a certain recovery threshold, denoted by $K(\vec{g},h)$, it can correct up to $\lfloor \frac{N-K(\vec{g},h)}{2} \rfloor$ errors.
With Lemma \ref{lemma:hamming}, proving Theorem 5 is equivalent to proving that the minimum recovery threshold of any scheme is $K$.

Suppose that a scheme $(\vec{g},h)$ is used for the computations. Then, we present the following lemma (Lemma 1 in [10]) which provides the converse bound of recovery threshold of computing any multilinear function $h$.
\begin{lemma}\label{lemma:lcc}
Given any multilinear function $h$, the recovery threshold $K(\vec{g},h)$ of any scheme $(\vec{g},h)$ satisfies
\begin{align}
    K(\vec{g},h) \geq \min\{(K-1) \textrm{deg} h+1, N - \lfloor N/K \rfloor +1 \}.
\end{align}
\end{lemma}
It is clear that the degree of function $h$ is at least $1$ since constant functions do not work in our problem. Moreover, the recovery threshold is a non-decreasing function on degree of $h$. By Lemma~\ref{lemma:lcc}, the recovery threshold $K(\vec{g},h)$ is lower bounded by $K$ which concludes the proof.